\documentclass[12pt]{umj_eng3}
\usepackage[english]{babel}
\usepackage{amsmath}
\usepackage{amssymb}
\usepackage{amsfonts, dsfont, graphicx, srcltx, color}

\newtheorem{theorem}{Theorem}[section]

\firstpage{125}

\renewcommand{\leq}{\leqslant}
\renewcommand{\geq}{\geqslant}

\allowdisplaybreaks
\numberwithin{equation}{section}

\begin{document}

 \thispagestyle{empty}

\title[Construction of exact solutions]{Construction of exact solutions \\ of nonlinear PDE via dressing chain in 3D}

\author{I.T. Habibullin, A.R. Khakimova}

\address{Ismagil Talgatovich Habibullin,
\newline\hphantom{iii} Institute of Mathematics,
 \newline\hphantom{iii} Ufa Federal Research Center, RAS
\newline\hphantom{iii} Chernyshevsky str.  112,
\newline\hphantom{iii} 450008, Ufa, Russia
}
\email{habibullinismagil@gmail.com}

\address{Aigul Rinatovna Khakimova,
\newline\hphantom{iii} Institute of Mathematics,
 \newline\hphantom{iii} Ufa Federal Research Center, RAS
\newline\hphantom{iii} Chernyshevsky str.  112,
\newline\hphantom{iii} 450008, Ufa, Russia
  }
\email{aigul.khakimova@mail.ru}

\thanks{\sc I.T. Habibullin, A.R. Khakimova, Construction of exact solutions of nonlinear PDE via dressing chain in 3D}
\thanks{\copyright \ Habibullin I.T., Khakimova A.R. \ 2024}
\thanks{\it Submitted September 20, 2024.}

\maketitle {\small
\begin{quote}
\noindent{\bf Abstract.} The duality between a class of the Davey~---~Stewartson type coupled systems and a class of two--dimensional Toda type lattices is discussed. A new coupled system related to the recently found lattice is presented. A method for eliminating nonlocalities in coupled systems by virtue of special finite reductions of the lattices is suggested. An original algorithm for constructing explicit solutions of the coupled systems based on the finite reduction of the corresponding lattice is proposed.  Some new solutions for coupled systems related to the Volterra lattice are presented as illustrative examples.

\medskip



\end{quote}
}

\bigskip
 
\section{Introduction}
In the articles \cite{LeznovShabatYamilov}, \cite{ShabatYamilov} a close connection has been discovered between integrable two--dimensional lattices and integrable partial differential equations in three independent variables. More precisely, the class of generalized symmetries of two--dimensional lattices contains a large variety of nonlinear integrable partial differential equations in 3D. In particular, this class contains such an important model as the Davey~---~Stewartson equation, which is related to the symmetries of the Toda lattice. After a such kind observation it was natural to expect that this duality would lead to the creation of new algorithms for finding particular solutions of coupled systems. However, this did not happen, since, as noted in the mentioned works, the presence of nonlocal variables creates severe problems for efficient use of dressing chains for constructing explicit solutions of coupled systems.

After the papers \cite{Habibullin13}--\cite{HabibullinKhakimova21} it became clear that the integrability of a two--dimensional chain can be fully revealed at the level of its finite--field reductions obtained by imposing a special type of truncating boundary conditions on the chain. The integrability criterion for a three-dimensional lattice is formulated as a requirement of Darboux integrability of the reduced systems.  This special type of boundary conditions for the chains has another remarkable property: it is compatible with arbitrary higher symmetry of the lattice under consideration. It immediately follows from this fact that special cut--off constraints preserve the duality between the lattices and the associated coupled systems. In other words, when the lattice is reduced to a finite--field  system, its symmetry becomes a symmetry for this reduction. An important step in this scheme consists of finding explicit expressions for the nonlocalities in terms of the local variables (see formulas \eqref{Vj}). Thus, at the reduction level, nonlocal variables can be completely eliminated.

Let us illustrate some of the notions considered below with an example. It was shown in \cite{ShabatYamilov} that the two--dimensional Volterra chain
\begin{equation}\label{Volterra}
u_{n,y}=u_n(v_{n+1}-v_n), \qquad v_{n,x}=v_n(u_n-u_{n-1})
\end{equation}
has a symmetry (coupled system) of the following form
\begin{equation}\label{csVolterra}
\begin{aligned}
&u_{n,t}=u_{n,xx}+\left(u_n^2+2u_nV_n\right)_x,\\
&v_{n,t}=-v_{n,xx}+\left(V^2_n\right)_y+(2u_nv_n)_x, \quad V_{n,y}=v_{n,x}.
\end{aligned}
\end{equation}
In other words, the relations
\begin{equation*}\label{definitionofsymmetry}
(u_{n,y})_t=(u_{n,t})_y, \qquad (v_{n,x})_t=(v_{n,t})_x
\end{equation*}
are satisfied identically if the derivatives with respect to $x$, $y$, and $t$ are replaced taking into account \eqref{Volterra} and \eqref{csVolterra}. For an arbitrary value of $n$, the relations \eqref{csVolterra} define a system of partial differential equations with independent variables $x$, $y$, and $t$, where $n$ is a hidden parameter. When moving from $n$ to $n+1$, the desired functions are calculated using an invertible B\"acklund transformation generated by the lattice \eqref{Volterra} (see \cite{ShabatYamilov})
\begin{equation}\label{Backlund1}
v_{n+1}=v_{n}+(\ln u_{n})_y,\qquad u_{n+1}=u_{n}+(\ln v_{n+1})_x, \qquad V_{n+1}=V_{n}+(\ln u_{n})_x.
\end{equation}

In this paper, we consider coupled systems corresponding to integrable lattices as the main object of study. The lattices rewritten as invertible B\"acklund transformations are interpreted as symmetries with discrete time $n$ for coupled systems. The aim of the work is to develop an algorithm for constructing particular solutions of partial differential equations with three independent variables using dressing chains.

\section{Examples of the coupled systems}

The Volterra chain admits a large class of symmetries (see \cite{ShabatYamilov}). For instance, one can easily derive from \eqref{csVolterra} another coupled system of the second order  
\begin{equation}\label{y-symm}
\begin{aligned}
&u_{n,\tau}=u_{n,yy}+\left( U^2_n\right)_x +(2u_nv_n)_y,\\
&v_{n,\tau}=-v_{n,yy}+\left(v^2_n+2v_n U_n\right)_y, \quad  U_{n,x}=u_{n,y},
\end{aligned}
\end{equation}
by using the involutions 
\begin{equation}\label{involut}
x\leftrightarrow -y, \quad t\leftrightarrow \tau, \quad u\leftrightarrow v,\quad U\leftrightarrow  V,\quad n \leftrightarrow -n.
\end{equation}
Its B\"acklund transformation is given by
\begin{equation*}
u_{n-1}=u_{n}-(\ln v_{n})_x, \qquad v_{n-1}=v_{n}-(\ln u_{n-1})_y, \qquad U_{n-1}=U_{n}-(\ln v_{n})_y.
\end{equation*}
By taking a linear combination of two symmetries given above we find a more complicated symmetry
\begin{equation}\label{csVolterra3}
\begin{aligned}
&u_{n,s}=\lambda u_{n,xx}+\mu u_{n,yy}-\lambda\left(u_n^2+2u_nV_n\right)_x-\mu\left( U_n^2\right)_x -\mu(2u_nv_n)_y,\\
&V_{n,y}=v_{n,x},\quad  \lambda\neq0,\\
&v_{n,s}=-\lambda v_{n,xx} -\mu v_{n,yy}-\lambda \left(V_n^2\right)_y-\lambda(2u_nv_n)_x -\mu\left(v_n^2+2v_n U_n\right)_y,\\
&U_{n,x}=u_{n,y}, \quad\mu\neq0.
\end{aligned}
\end{equation}

A large class of the Toda type integrable lattices is presented in \cite{ShabatYamilov} where the related coupled systems are given as well. One extra lattice of this type 
\begin{equation}\label{newlattice}
\begin{aligned}
&u_{n,xy} = \alpha_n(u_{n,x} - u^2_n - 1)(u_{n,y} - u^2_n - 1) + 2 u_n(u_{n,x}+u_{n,y}-u^2_n - 1),\\
&\alpha_n = \frac{1}{u_n - u_{n-1}} - \frac{1}{u_{n+1}-u_n}
\end{aligned}
\end{equation}
has recently been found in \cite{HabibullinKuznetsova20}.
Symmetries of lattice \eqref{newlattice} in the directions of $x$ and $y$ are as follows (see \cite{HabibullinKhakimova24}):
\begin{equation}\label{shortx1}
\begin{aligned}
&u_{n,t}=u_{n,xx}-2u_nu_{n,x}+u_n^2+1-2(u_n^2-u_{n,x}+1)H_n, \\
&H_n=(T-1)^{-1}D_{x}\log \frac{u_{n,x}-u^2_n-1}{u_{n+1}-u_n},\\
&D_{y}H_{n}=- D_{x} \frac{u_{n,y}-u_nu_{n-1}-1}{u_{n}-u_{n-1}}
\end{aligned}
\end{equation}
and respectively:
\begin{equation}\label{shorty1}
\begin{aligned}
&u_{n,\tau}=u_{n,yy}-2u_nu_{n,y}+u_n^2+1-2(u_n^2-u_{n,y}+1)Q_n,\\
&Q_n=(T-1)^{-1}D_{y}\log \frac{u_{n+1,y}-u^2_{n+1}-1}{u_{n+1}-u_{n}},\\
&D_xQ_{n-1}=D_y \frac{u_{n-1,x}-u_{n}u_{n-1}-1}{u_{n}-u_{n-1}}.
\end{aligned}
\end{equation}

Note that the symmetries \eqref{shortx1} and \eqref{shorty1} depend significantly on the discrete parameter $n$, since they contain variables with shifted arguments.

The corresponding coupled systems of the lattice \eqref{newlattice} obtained from the second--order symmetries \eqref{shortx1} and \eqref{shorty1} have the form:
\begin{equation}\label{csx7}
\begin{aligned}
&u_{t}=u_{xx}-2uu_{x}+u^2+1-2(u^2-u_{x}+1)H,\\
&v_{t}=-v_{xx}  +2(v_{x}-v^2-1)H-\frac{2v_{x}^2}{u-v}\\
&\quad\quad+\frac{ 2(v_{x}-v^2-1)u_{x}}{u-v}+\frac{2(uv+1)v_{x}}{u-v}+v^2+1,\\
&D_{y}H=- D_{x} \frac{u_{y}-uv-1}{u-v}
\end{aligned} 
\end{equation}
and
\begin{equation}\label{csy8}
\begin{aligned}
&u_{\tau}= -u_{yy}+2(u_{y}-u^2-1)Q+\frac{2u^2_{y}}{u-v}\\
&\quad \quad -\frac{2(u_{y}-u^2-1)v_{y}}{u-v}-\frac{2(uv+1)u_{y}}{u-v}+u^2+1,\\
&v_{\tau}=v_{yy}-2vv_{y}+v^2+1+2(v_{y}-v^2-1)Q,\\
&D_xQ=D_y\left(\frac{v_x-uv-1}{u-v}\right),
\end{aligned} 
\end{equation}
where $u:=u_n$ and $v:=u_{n-1}$. Obviously systems \eqref{csx7} and \eqref{csy8} do not contain any variables $u$ and $v$ with shifted values of $n$.

The lattice \eqref{newlattice}, supplemented by the equation for the nonlocality  $H_n$, defines an invertible B\"acklund transformation 
\begin{align*}
&v_{n-1}=v_n-\frac{(u_n-v_n)(v^2_n-v_{n,x}+1)(v^2_n-v_{n,y}+1)}{(u_n-v_n)\left(v_{n,xy}-2v_n(v_{n,x}+v_{n,y}-v^2_n-1)\right)+(v^2_n-v_{n,x}+1)(v^2_n-v_{n,y}+1)},\\
&u_{n-1}=v_n,\\
&H_{n-1}=H_{n} - D_x\log \frac{v_{n,x}-v_{n}^2-1}{u_{n}-v_{n}}
\end{align*}
for the coupled system (\ref{csx7}). In a similar way one can derive the B\"acklund transformation for coupled system (\ref{csy8}).

\section{Finite reductions of 3D lattices  compatible with symmetries}

It is shown in our articles \cite{Habibullin13}--\cite{Kuznetsova19} that each known integrable Toda type lattice admits cut--off boundary conditions allowing to reduce the lattice to a hierarchy of hyperbolic systems integrable in sense of Darboux. Apparently, these boundary conditions are compatible with a large class of higher symmetries of the lattice. We say that a truncation boundary condition for a lattice is compatible with its symmetry if the truncation preserves the commutativity property of the lattice and the symmetry \cite{GurelHabibullin97}. Below in this section we discuss how to get rid of the nonlocalities that arise within the symmetry approach. To do this, we pass from chains to their finite--dimensional reductions, then use conservation laws to express the nonlocalities in terms of dynamical variables.


Let us concentrate on finite--field reductions of the Volterra chain obtained by imposing the following type truncation conditions $u_{-k}=0$ and $v_{m+k}=0$, $k=1,2,3,\ldots$:
\begin{equation}\label{*1}
\begin{aligned}
&u_{0,y}=u_0(v_1-v_0), \qquad && v_{0,x}=v_0u_0,\\
&u_{1,y}=u_1(v_2-v_1), \qquad && v_{1,x}=v_1(u_1-u_0),\\
&\ldots \qquad && \ldots \\
&u_{m,y}=-u_mv_m, \qquad && v_{m,x}=v_m(u_{m}-u_{m-1}).
\end{aligned}
\end{equation}
By summing consecutively equations of the obtained system we arrive at an equation $\sum^{m}_{i=0}\left(u_{i,y}+v_{i,x}\right)=0$ that can be represented as a conservation law
\begin{align*}
\frac{\partial}{\partial y}\sum^{m}_{i=0}u_{i}  +   \frac{\partial}{\partial x}\sum^{m}_{i=0} v_{i}=0.
\end{align*}
Replacing $v_{i,x}=V_{n,y}$ due to the definition of the nonlocal variables (see \eqref{csVolterra}) we find that expression 
$\sum^{m}_{i=0}\left(u_{i}+V_{i}\right)$ 
does not depend on $y$:
\begin{equation*}\label{*2}
\frac{\partial}{\partial y}\sum^{m}_{i=0}\left(u_{i}+V_{i}\right)=0.
\end{equation*}
Therefore if we require that the relation
\begin{align}\label{*3}
\sum^{m}_{i=0}V_{i}=-\sum^{m}_{i=0}u_{i}
\end{align}
is satisfied for some fixed value $y=y_0$ then it would hold for all $y$. Due to the B\"acklund transformation \eqref{Backlund1} we have the first order linear discrete equation $V_{n+1}=V_n+h_n$, where $h_n=(\ln u_n)_x$. The equation is easily solved 
\begin{align}\label{*4}
V_{j}=V_{0}+\sum^{j-1}_{i=0}h_{i}.
\end{align}
By using the explicit solution \eqref{*4} one can derive a useful formula
\begin{align}\label{*5}
\sum^{m}_{k=1}V_{k}=mV_{0}+\sum^{m}_{j=1}\sum^{j-1}_{i=0}h_{i}.
\end{align}
Now we rewrite \eqref{*3} in the following form 
\begin{align}\label{*6}
\sum^{m}_{k=1}V_{k}=-V_{0}-\sum^{m}_{i=0}u_{i}.
\end{align}
Comparison of \eqref{*5} and \eqref{*6} leads to an explicit expressions for all of the variables $V_j$, $j=\overline{0,m}$ in terms of the dynamical variables $u_0$, \ldots, $u_m$ and their first order derivatives with respect to~$x$: 
\begin{equation}\label{Vj}
\begin{aligned}
&V_{0}=-\frac{1}{m+1}\left(\sum^{m}_{j=1}\sum^{j-1}_{i=0}h_{i} + \sum^{m}_{i=0}u_{i}\right),\\
&V_{j}=-\frac{1}{m+1}\left(\sum^{m}_{j=1}\sum^{j-1}_{i=0}h_{i} + \sum^{m}_{i=0}u_{i}\right)+\sum^{j-1}_{i=0}h_{i}, \quad j=\overline{0,m}.
\end{aligned}
\end{equation}

As a result we determine a usual second order symmetry (without nonlocalities) of the hyperbolic system \eqref{*1}:
\begin{equation}\label{*8}
\begin{aligned}
&u_{n,t}=u_{n,xx}+\left(u_n^2+2u_nV_n\right)_x, \quad \mbox{for} \quad n=\overline{0,m},\\
&v_{n,t}=-v_{n,xx}+\left(V^2_n\right)_y+(2u_nv_n)_x,
\end{aligned}
\end{equation}
where functions $V_j=F_j([u])$ are found due to \eqref{Vj}. In a similar way one can study the nonlocality $U_n$ of the coupled system \eqref{y-symm} corresponding to the Volterra lattice. The explicit expressions for the variables $U_0,\,U_1,\,\dots,U_m$ are easily derived from \eqref{Vj} due to involutions~\eqref{involut}.

\textbf{Example 1.} For $m=0$ formula \eqref{Vj} takes the form $V_0=-u_0$ (see \eqref{nonlocalityV}, below). It is considered in section 4, where explicit solution for the coupled system is constructed. 

\textbf{Example 2.} If $m=1$ the nonlocalities are given by
\begin{equation*}\label{V0V1}
\begin{aligned}
&V_{0}=-\frac{1}{2}\left(h_0+u_0+u_1\right), \quad h_0=(\ln u_0)_x,\\
&V_{1}=-\frac{1}{2}\left(h_0+u_0+u_1\right)+h_0.
\end{aligned}
\end{equation*}
The corresponding symmetry takes the form
\begin{equation}\label{symm1}
\begin{aligned}
&u_{0,t}=-u_{0,x}u_1-u_0u_{1,x}, \\
& v_{0,t}=-u_0v_0u_1,\\
&u_{1,t}=u_{1,xx}+\frac{u_1}{u_0}u_{0,xx}+\frac{\left(u_{0,x}-u_0^2\right)u_{1,x}}{u_0}-\frac{\left(u_{0,x}+u_0^2\right)u_1u_{0,x}}{u_0^2}, \\
& v_{1,t}=\frac{u_1v_1u_{0,x}}{u_0}+u_{1,x}v_1.
\end{aligned}
\end{equation}
It is easily verified that \eqref{symm1} is really a symmetry to the reduced system:
\begin{equation*}\label{sysm1}
\begin{aligned}
&u_{0,y}=u_0(v_1-v_0), \qquad && v_{0,x}=v_0u_0,\\
&u_{1,y}=-u_1v_1, \qquad && v_{1,x}=v_1(u_{1}-u_{0}).
\end{aligned}
\end{equation*}

Recall that system \eqref{*1} is integrable in sense of Darboux, i.e. it admits a complete set of characteristic integrals. It is known that symmetries  of the Darboux integrable systems are linearized (see, for instance, \cite{Zhiber2001}). Therefore \eqref{*8} is linearized by an appropriately chosen differential substitution.

\section{Construction of exact solutions to the coupled systems via integrable reductions of the dressing chains}

Now we discuss how the dressing chain can be used to construct explicit solutions to the coupled systems. As an illustrative example we take the system 
\begin{equation}\label{csVolterraR}
\begin{aligned}
&u_{t}=u_{xx}+\left(u^2+2uV\right)_x,\\
&v_{t}=-v_{xx}+\left(V^2\right)_y+(2uv)_x, \quad V_y=v_x,
\end{aligned}
\end{equation}
corresponding to the Volterra lattice (here $u:=u_n$ and $v:=v_n$)
\begin{equation*}
u_{n,y}=u_n(v_{n+1}-v_n), \qquad v_{n,x}=v_n(u_n-u_{n-1}).
\end{equation*}
Let us consider its reduction
\begin{equation}\label{reduction2}
u_{y}=-uv, \qquad   v_{x}=uv      
\end{equation}
obtained due to cutting--off constraint $u_{-k}=0$, $v_{k}=0$, $k=1,2,3,\ldots$. Here the sought functions are $u:=u_0$ and $v:=v_0$. The functions 
$$I=\frac{u_x}{u}-u \quad  \mbox{and} \quad J=\frac{v_y}{v}+v$$
are characteristic integrals of the system. Indeed, it is checked straightforwardly that the necessary conditions $D_yI=0$ and $D_xJ=0$ for the integrals definitely hold. Therefore we have a system of differential equations (Bernoulli equations)
\begin{equation*}\label{ODE}
\frac{u_x}{u}-u=f_1(x), \qquad   \frac{v_y}{v}+v=f_2(y)      
\end{equation*} 
for searching solution to system \eqref{reduction2}, where $f_1$ and $f_2$ are arbitrary functions.

It is easy to check that general solution of the system can be parametrized in the following form 
\begin{equation}\label{solution}
u(x,y)=-\frac{W_x}{W}=\frac{\rho'(x)}{\varphi(y)-\rho(x)}, \qquad  v(x,y)=\frac{W_y}{W}=\frac{-\varphi'(y)}{\varphi(y)-\rho(x)},      
\end{equation}
where $W=\varphi(y)-\rho(x)$ and $W_x$, $W_y$ denote the derivatives of $W$ with respect to $x$ and $y$. Here the functions $\varphi(y)$, $\rho(x)$ are chosen arbitrarily. Note that $W_{xy}=0.$

Now we assume that functions $u(x,y)$ and $v(x,y)$ depend on one more independent variable $t$ due to system \eqref{csVolterraR}. In other words we have $\varphi=\varphi(y,t)$ and $\rho=\rho(x,t)$. Then by integrating equation $V_y=v_x$ we derive an explicit expression for the nonlocality
\begin{equation*}\label{nonlokV}
V=-\int \frac{\varphi_y\rho_xdy}{W^2}=\frac{\rho_x}{W}+R(x,t)=-u+R(x,t).
\end{equation*}
Let us set $R(x,t)=0$ for simplicity. Then we get (see also Example 1)
\begin{equation}\label{nonlocalityV}
V=-u.
\end{equation}
Afterward the coupled system \eqref{csVolterra} turns into 
\begin{equation}\label{csFinal}
u_t=u_{xx}-2uu_x, \qquad v_t=-v_{xx}+2uu_y+2(uv)_x.
\end{equation}
 Using the substitution \eqref{solution}, we reduce the system \eqref{csFinal} to an overdetermined system of equations with a single sought function $W$.
To apply the substitution \eqref{solution} we have to use the explicit representations of the derivatives of $u$ and $v$:
\begin{equation*}\label{explicitderivatives}
\begin{aligned}
&u_y=-v_x=\frac{W_x W_y}{W^2},\qquad &&u_x=-\frac{W_{xx}}{W}+\frac{W_x W_y}{W^2}, \\
&u_t=-\frac{W_{xt}}{W}+\frac{W_x W_t}{W^2}, \qquad  &&v_t=\frac{W_{yt}}{W}-\frac{W_y W_t}{W^2},\\
&v_{xx}=-\frac{W_{xx}W_y}{W^2}+2\frac{W_x^2W_y}{W^3}, \qquad &&u_{xx}=-\frac{W_{xxx}}{W}+3\frac{W_x W_{xx}}{W^2}-2\frac{W_x^3}{W^3}.
\end{aligned}
\end{equation*}

The formulas above allow us to bring system \eqref{csFinal} to the following form
\begin{equation}\label{Usystem}
\begin{aligned}
&WW_{ty}=W_yW_{t}-W_{xx}W_y,  \\
&WW_{tx}=WW_{xxx}+W_xW_{t}-W_xW_{xx}.
\end{aligned}
\end{equation}
The next step is to solve it explicitly. Let us start with the second equation of \eqref{Usystem}. First, we represent the equation as
\begin{equation*}
\frac{W W_{xt}-W_x W_t}{W^2}=\frac{WW_{xxx}-W_xW_{xx}}{W^2}.
\end{equation*}
Then integrating the latter we get
\begin{equation*}
\frac{W_t}{W}=\frac{W_{xx}}{W}+g(t,y).
\end{equation*}

Now we simplify the first equation in \eqref{Usystem}     due to the relation 
\begin{equation}\label{d2}
W_t=W_{xx}+g(t,y)W
\end{equation}
and obtain an equation of the form
\begin{equation}\label{d3}
W_{yt}=g(t,y)W_y.
\end{equation}
If we apply the operator $D_y$ of the total differentiation with respect to $y$ to both sides of \eqref{d2} and then simplify it in virtue of equation $W_{xxy}=0$ we arrive at the relation
\begin{equation}\label{d4}
W_{yt}=g(t,y)W_y+g_y(t,y)W.
\end{equation}
Comparing relations \eqref{d3} and \eqref{d4} gives rise to $g_y(t,y)=0$ or, the same,
\begin{equation*}\label{d5}
g(t,y)=g(t).
\end{equation*}

Analyzing the reasoning above we can conclude that the desired function $W=W(x,y,t)$ is a solution to the system
\begin{equation}\label{finalsyst}
\begin{aligned}
&W_{xy}=0,  \\
&W_{t}=W_{xx}+ g(t)W, \\
&W_{ty}=g(t)W_y.
\end{aligned}
\end{equation}
Obviously the third equation of the system is easily integrated, since it is of the form 
$$\frac{\partial}{\partial t}\ln W_y=g(t).$$ 
Hence it implies 
$$\ln W_y= \ln G(t) +\ln F_1(y),$$ 
where $\ln G(t)=\int_0^t g(\tau)d\tau$ and the constant of integration $\ln F_1(y)$ does not depend on $x$ due to the first equation in \eqref{finalsyst}. Then we integrate the obtained equation $W_y=G(t)F_1(y)$ with respect to $y$. It is convenient to present the result in the form   
\begin{equation}\label{d6}
W=G(t)(F(y)+S(x,t)),
\end{equation}
where $F(y)=\int_0^y F_1(z)dz$.

No we substitute \eqref{d6} into the second equation of \eqref{finalsyst}. After a slight simplification we obtain the heat equation for $S(x,t)$
\begin{equation}\label{d7}
S_t=S_{xx}.
\end{equation}
Therefore general solution to the system \eqref{finalsyst} is given by \eqref{d6} with arbitrary $G(t)$ and $F(y)$ and with an arbitrary solution $S(x,t)$ to \eqref{d7}. In other words we have 

\begin{theorem}\label{th} Assume that $S(x,t)$ is a solution to the equation \eqref{d7} and $F(y)$ is an arbitrary smooth function, then functions defined due to the rule 
\begin{equation}\label{d8}
\begin{aligned}
&u(x,y,t)=-\frac{\partial}{\partial x} \ln (S(x,t)+F(y)),\\
&v(x,y,t)=\frac{\partial}{\partial y} \ln (S(x,t)+F(y)),\\
&V(x,y,t)=\frac{\partial}{\partial x} \ln (S(x,t)+F(y))
\end{aligned}
\end{equation}
give a solution to the coupled system \eqref{csVolterra}.
\end{theorem}

The statement of the theorem \ref{th} is easily verified by a simple substitution.

As is known, the solution of the heat conduction equation \eqref{d7} is given in a closed form according to the Poisson formula 
\begin{equation*}\label{Poisson}
S(x,t)=\frac{1}{2\sqrt{\pi t}}\int_{R}S_0(\xi)e^{-\frac{(x-\xi)^2}{4t}}d\xi,
\end{equation*}
where $\left.S\right|_{t=0}=S_0(x)$ is a continuous and bounded function. Therefore solution \eqref{d8} of coupled system \eqref{csVolterra} depends on two arbitrary functions $S_0(x)$ and $F(y)$.

\section{The second example}
Taking a linear combination of two symmetries, we find a coupled system that depends symmetrically on $x$ and $y$ (see \eqref{csVolterra3} above)
\begin{equation}\label{csVolterra30}
\begin{aligned}
&u_{n,s}=\lambda u_{n,xx}+ \mu u_{n,yy}+\lambda\left(u_n^2+2u_nV_n\right)_x +\mu\left( U_n^2\right)_x +\mu(2u_nv_n)_y,\\
&V_{n,y}=v_{n,x},\quad  \lambda\neq0,\\
&v_{n,s}=-\lambda v_{n,xx} -\mu v_{n,yy}+\lambda \left(V_n^2\right)_y+\lambda(2u_nv_n)_x +\mu\left(v_n^2+2v_n U_n\right)_y,\\
&U_{n,x}=u_{n,y}, \quad\mu\neq0.
\end{aligned}
\end{equation}
The boundary conditions $u_{-k}=0$ and $v_{k}=0$, $k=1,2,3,\ldots$ imposed on the Volterra chain are compatible with all symmetries.
Therefore to construct solutions of system \eqref{csVolterra3} one can use the same ansatz
\begin{equation*}\label{solution2}
u(x,y)=-\frac{W_x}{W}, \qquad  v(x,y)=\frac{W_y}{W}      
\end{equation*}
as in the previous example. 
Here $W=\varphi(y)-\rho(x)$ with arbitrary functions $\varphi(y)$, $\rho(x)$ and $W_x$, $W_y$ denote the derivatives of $W$ with respect to $x$ and $y$. We choose the nonlocalities as $V=-u$ and $U=-v$.

As a result, we arrive at the following system of equations:
\begin{align*}
W_{sx}-&\lambda W_{xxx}- \mu W_{xyy}-\frac{W_xW_s}{W}      +\lambda\frac{W_{xx}W_x}{W}    +2\mu\frac{W_{xy}W_y}{W}-\mu\frac{W_{x}W_{yy}}{W}=0,\\
W_{sy}+&\lambda W_{xxy}+ \mu W_{yyy}-\frac{W_yW_s}{W}      -\mu\frac{W_{yy}W_y}{W}    -2\lambda\frac{W_{xy}W_x}{W}+\lambda\frac{W_{y}W_{xx}}{W}=0.
\end{align*}
We integrate the first equation with respect to $x$ and the second one integrate with respect to~$y$.  Finally we get one and the same equation
\begin{equation}\label{csVolterra5}
W_{s}-\lambda W_{xx}+ \mu W_{yy}-g(s)W=0.    
\end{equation}
Since function $W$ has the form $W=\varphi(y,s)-\rho(x,s)$, then from the equation \eqref{csVolterra5} it follows
\begin{equation*}
\varphi_{s}-\rho_{s}+\lambda \rho_{xx}+ \mu \varphi_{yy}-g(s)\left(\varphi-\rho\right)=0.    
\end{equation*}
Separating the variables in the last equation we get
\begin{equation*}
\varphi_{s}+ \mu \varphi_{yy}-g(s)\varphi=\rho_{s}-\lambda \rho_{xx}-g(s)\rho=:\gamma(s)  
\end{equation*}
Therefore we have
\begin{equation}\label{heat0}
\begin{aligned}
&\varphi_{s}+ \mu \varphi_{yy}-g(s)\varphi=\gamma(s),\\
&\rho_{s}-\lambda \rho_{xx}-g(s)\rho=\gamma(s).
\end{aligned}
\end{equation}
Let us simplify the obtained equations by means of linear transformations 
$$\varphi(y,s)=\beta(s)\psi(y,s)+\alpha(s), \qquad \rho(x,s)=\beta(s)\theta(x,s)+\alpha(s),$$ 
where $\beta(s)$ and $\alpha(s)$ are solutions of the following equations $\beta'-\beta g=0$, $\alpha'-\alpha g=\gamma$. As a result, equations \eqref{heat0} take the form
\begin{equation}\label{heat}
\psi_{s}+\mu \psi_{yy}=0, \qquad  \theta_{s}-\lambda \theta_{xx}=0.     
\end{equation}

\begin{theorem} Assume that $\psi(y,s)$ and $\theta(x,s)$ are arbitrary solutions to the equations \eqref{heat}. Then the functions given in the following way
\begin{equation}\label{exactsolution}
\begin{aligned}
&u=\frac{\theta_x}{\psi-\theta},  \qquad
&&v=\frac{\psi_y}{\psi-\theta}, \\
&U=-\frac{\psi_y}{\psi-\theta},\qquad
&&V=-\frac{\theta_x}{\psi-\theta}
\end{aligned}
\end{equation}
define a solution to the coupled system \eqref{csVolterra30}.
\end{theorem}

Note that the solution of the heat equation \eqref{heat} for $\lambda=-\mu=1$ is explicitly expressed through the Poisson integral. Consequently, it is easy to write down formulas expressing the solution of the system \eqref{csVolterra30} containing two arbitrary functions.

\section{Construction of a particular solution of lattice (1.8)} 

In this section we construct a particular solution of the lattice \eqref{newlattice}, subject to the following additional constraints
\begin{align}\label{set}
u:=u_0, \quad \mbox{for } n\geq 1, \qquad u_{n}=(-1)^{n+1}i, \quad \mbox{for } n\leq -1, \qquad u_{n}=(-1)^{n}i.
\end{align}
Then the lattice \eqref{newlattice} takes the form
\begin{equation*}\label{lattice0}
u_{xy}=\frac{2uu_{x}u_{y}}{u^2+1}.
\end{equation*}
Next, we use the integrals of lattice \eqref{newlattice} found in \cite{HabSakieva24}, which for the choice of $u_{n}$ by virtue of \eqref{set} have the form:
\begin{equation*}\label{Integrals}
J=\frac{u_{x}}{u^2+1}, \qquad I=\frac{u_{y}}{u^2+1}.
\end{equation*}
Recall that a function $J(u,u_{x},u_{xx},u_{xxx},\ldots)$ is called a $y$--integral if the condition 
$$D_yJ(u,u_{x},u_{xx},u_{xxx},\ldots)=0$$ 
is satisfied. The $x$--integral is defined similarly. 

From the condition $D_yJ(u,u_{x},u_{xx},u_{xxx},\ldots)=0$ we find: 
$$J(u,u_{x},u_{xx},u_{xxx},\ldots)=f(x)$$
or in our case we have:
$$ \frac{u_{x}}{u^2+1}=f(x). $$
Integrating the last expression, we obtain:
$$u=\tan (F(x)+G(y)),$$
where $F'(x)=f(x)$. Thus $F(x)$ and $G(y)$ are arbitrary functions.

Let us rewrite the found solution in general form
\begin{equation}\label{solution0}
u=\tan W(x,y),
\end{equation}
where $W(x,y)=F(x)+G(y)$.

Now we substitute \eqref{solution0} into the symmetry \eqref{csx7}:
\begin{align*}
u_{n,x_2}=u_{n,x_1x_1}-2u_nu_{n,x_1}+u_n^2+1-2(u_n^2-u_{n,x_1}+1)H_n
\end{align*}
of equation \eqref{newlattice}. First, we simplify this symmetry due to the restrictions \eqref{set}, namely, we find out what the value of nonlocality $H$. Let us put $x:=x_1$, $t:=x_2$, $y:=y_1$, $H:=H_0$:
\begin{align}
& u_{t}=u_{xx}-2uu_{x}+u^2+1-2(u^2-u_{x}+1)H, \label{symm0}\\
& D_{y}H=- D_{x} \left(\frac{u_{y}+iu-1}{u+i}\right). \nonumber
\end{align}
We integrate the last equality with respect to $y$ and find:
\begin{align*}
H=-\frac{u_x}{u+i}-\varphi(x),
\end{align*}
where $\varphi(x)$ is arbitrary function. Here we will restrict ourselves to considering case $\varphi(x)=0$. Due to this condition symmetry \eqref{symm0} can be written as:
\begin{align}\label{symm00}
u_t=u_{xx}-\frac{2u_x^2}{u+i}-2iu_x+u^2+1.
\end{align}
Let us substitute $u=\tan W(x,y,t)$ into \eqref{symm00} and find:
\begin{align*}
W_t=W_{xx}+i\left(2W_x^2-2W_x-i\right).
\end{align*}
Now we differentiate the last equation with respect to $x$:
\begin{align*}
W_{t,x}=W_{xxx}+i\left(4W_xW_{xx}-2W_{xx}\right)
\end{align*}
and make the substitution $W_x=\frac{1}{2}\left(-i\tilde{W}+1\right)$. Then the equation will be reduced to the Burgers equation:
\begin{align*}
\tilde{W}_t=\tilde{W}_{xx}+2\tilde{W}\tilde{W}_x.
\end{align*}
As is known, the Burgers equation is reduced to the heat equation
\begin{align*}
\bar{W}_t=\bar{W}_{xx}
\end{align*}
by means of the Cole~---~Hopf substitution $\tilde{W}=- \frac{\bar{W}_x}{\bar{W}}$.
Therefore the sought particular solution can be written in the form
\begin{align*}
u&=\tan(W)=\tan\left(D_x^{-1}\left(\frac{1}{2}\left(-i\tilde{W}+1\right)\right)\right)\\
&=\tan\left(D_x^{-1}\left(\frac{1}{2}\left(i\frac{\bar{W}_x}{\bar{W}}+1\right)\right)\right)\\
&=\tan\left(\frac{i}{2}\ln(\bar{W}(x,y,t))+\frac{x}{2}+C(y,t)\right),
\end{align*}
where $\bar{W}(x,y,t)$ is a solution of the heat equation. 
Let us determine the dependence of function $\bar{W}(x,y,t)$ on variable $y$. To do this, we substitute the found solution into the $x$--integral $\frac{u_{y}}{u^2+1}=g(y,t)$, where $g(y,t)$ is an arbitrary function. After simplification we obtain:
\begin{align*}
\bar{W}(x,y,t)=\bar{F}(x,t)\bar{G}(y,t), 
\end{align*}
where $\bar{F}(x,t)$ is arbitrary as well, $\bar{G}(y,t)=e^{2iC(y,t)-2i\int g(y,t)dy}$.
Finally, the particular solution of lattice \eqref{newlattice} takes the form:
\begin{align*}
u=\tan\left(\frac{i}{2}\ln(\bar{F}(x,t)\bar{G}(y,t))+\frac{x}{2}+C(y,t)\right).
\end{align*}

Turning to the corresponding coupled system \eqref{csx7} we can convince that its solution is given~by  
\begin{align*}
u&=\tan\left(\frac{i}{2}\ln(\bar{F}(x,t)\bar{G}(y,t))+\frac{x}{2}+C(y,t)\right),\\
v&=-i.
\end{align*}

\section*{Conclusion}

The problem of constructing explicit solutions for multidimensional integrable models is studied by many authors: Shabat, Zakharov, Novikov, Krichever, Manakov, Grinevich, Santini, Fokas, Taimanov, Konopelcnenko, Bogdanov, Ferapontov, Pavlov, Dryuma and others. A great variety of tools were suggested (see, for instance, \cite{GrinevichSantini}--\cite{KHKh23}).   

Here we discussed the dressing chains method that provides an effective tool for constructing explicit solutions for integrable nonlinear PDE in the dimension 1+1 (see, for instance, \cite{ShabatYamilov91}, \cite{Smirnov22} and references therein). However in 3D some difficulties arise due to the  nonlocal variables (see \cite{LeznovShabatYamilov}). Examples considered in the article convince  that to overcome these difficulties one can use finite reductions of the dressing chains obtained by imposing cutting off constraints preserving integrability. Besides the degenerate boundary conditions related to reductions integrable in the sense of Darboux one can use also more general boundary conditions compatible with the integrability property of the lattices (see \cite{GurelHabibullin97}).


\begin{thebibliography}{35}  

\bibitem{LeznovShabatYamilov} A.N. Leznov, A.B. Shabat, R.I. Yamilov. {\it Canonical transformations generated by shifts in nonlinear lattices}~// Phys. Lett., A \textbf{174}:5--6, 397--402 (1993).

\bibitem{ShabatYamilov} A.B. Shabat, R.I. Yamilov. {\it To a transformation theory of two--dimensional integrable systems}~// Phys. Lett., A, \textbf{227}:1--2, 15--23 (1997). 

\bibitem{Habibullin13} I.T. Habibullin. {\it Characteristic Lie rings, finitely--generated modules and integrability conditions for (2+1)--dimensional lattices}~// Phys. Scr. \textbf{87}:6, 065005 (2013).

\bibitem{HabibullinPoptsova18} M.N. Poptsova, I.T. Habibullin. {\it Algebraic properties of quasilinear two--dimensional lattices connected with integrability}~// Ufa Math. J. \textbf{10}:3, 86--105 (2018).

\bibitem{HabibullinKuznetsova20} I.T. Habibullin, M.N. Kuznetsova. {\it A classification algorithm for integrable two--dimensional lattices via Lie~---~Rinehart algebras}~// Theor. Math. Phys. \textbf{203}:1, 569--581 (2020).

\bibitem{Kuznetsova19} M.N. Kuznetsova. {\it Classification of a subclass of quasilinear two--dimensional lattices by means of characteristic algebras}~//  Ufa Math. J. \textbf{11}:3, 10--131 (2019).

\bibitem{HabibullinKhakimova21} I.T. Habibullin, A.R. Khakimova. {\it Characteristic Lie algebras of integrable differential--difference equations in 3D}~// J. Phys. A, Math. Theor. \textbf{54}:29, 295202, 34 p. (2021).

\bibitem{HabibullinKhakimova24} I.T. Habibullin, A.R. Khakimova. {\it Symmetries of Toda type 3D lattices}~// Preprint, arXiv:2409.07017 [nlin.SI] (2024).

\bibitem{GurelHabibullin97} B. G\"urel, I. Habibullin. {\it Boundary conditions for two--dimensional integrable chains}~// Phys. Lett., A \textbf{233}:1, 68--72 (1997).

\bibitem{Zhiber2001} A.V. Zhiber, V.V. Sokolov. {\it Exactly integrable hyperbolic equations of Liouville type}~// Russ. Math. Surv. \textbf{56}:1, 61–101 (2001).

\bibitem{HabSakieva24} I.T. Habibullin, A.U. Sakieva. {\it On integrable reductions of two--dimensional Toda--type lattices}~// Part. Diff. Eq. Appl. Math. \textbf{11} 100854 (2024). 

\bibitem{ShabatYamilov91} A.B. Shabat, R.I. Yamilov. {\it Symmetries of nonlinear chains}~// Leningr. Math. J. \textbf{2}:2, 377--400 (1991).

\bibitem{Smirnov22} V.B. Matveev, A.O. Smirnov. {\it Dubrovin method and Toda lattice}~// St. Petersbg. Math. J. \textbf{34}:6, 1019--1037 (2023).


\bibitem{GrinevichSantini} P.G. Grinevich, P.M. Santini. {\it The finite--gap method and the periodic Cauchy problem for (2+1)--dimensional anomalous waves for the focusing Davey~---~Stewartson 2 equation}~// Russ. Math. Surv. \textbf{77}:6, 1029--1059 (2022).

\bibitem{Taimanov21}  I.A. Taimanov. {\it The moutard transformation for the Davey~---~Stewartson II equation and its geometrical meaning}~// Math. Notes \textbf{110}:5, 754--766 (2021). 

\bibitem{Zakharov85} V.E. Zakharov, S.V. Manakov. {\it Construction of higher--dimensional nonlinear integrable systems and of their solutions}~// Funct. Anal. Appl. \textbf{19}:2, 89--101 (1985).

\bibitem{KulaevShabat} R.Ch. Kulaev, A.B. Shabat. {\it Darboux system and separation of variables in the Goursat problem for a third order equation in $R^3$}~// Russ. Math. \textbf{64}:4, 35--43 (2020).

\bibitem{Konopelchenko} B.G. Konopelchenko, B.T. Matkarimov. {\it Inverse spectral transform for the nonlinear evolution equation generating the Davey~---~Stewartson and Ishimori equations}~// Stud. Appl. Math. \textbf{82}:4, 319--359 (1990).

\bibitem{Kuznetsova23} M.N. Kuznetsova. {\it Construction of localized particular solutions of chains with three independent variables}~// Theor. Math. Phys. \textbf{216}:2, 1158--1167 (2023).

\bibitem{KHKh23} M.N. Kuznetsova, I.T. Habibullin, A.R. Khakimova. {\it On the problem of classifying integrable chains with three independent variables}~// Theor. Math. Phys. \textbf{215}:2, 667--690 (2023).

\end{thebibliography}
\end{document}